\documentclass[
    ,final            
  ]
  {aipproc}

\layoutstyle{6x9}

\usepackage{amsmath}
\usepackage{bm}

\def\centeron#1#2{{\setbox0=\hbox{#1}\setbox1=\hbox{#2}\ifdim
\wd1>\wd0\kern.5\wd1\kern-.5\wd0\fi
\copy0\kern-.5\wd0\kern-.5\wd1\copy1\ifdim\wd0>\wd1
\kern.5\wd0\kern-.5\wd1\fi}}
\def\lsim{\;\centeron{\raise.35ex\hbox{$<$}}{\lower.65ex\hbox
{$\sim$}}\;}                                                 
\def\gsim{\;\centeron{\raise.35ex\hbox{$>$}}{\lower.65ex\hbox
{$\sim$}}\;}

\begin{document}

\title{Spin Correlations and Velocity-Scaling in NRQCD 
Matrix Elements
\protect\footnote{Talk presented by G.~T.~Bodwin.}
}


\author{Geoffrey T.~Bodwin}{
  address={HEP Division, Argonne National Laboratory, 9700 South Cass 
Avenue, Argonne, IL 60439}
}

\author{Jungil Lee}{
  address={Department of Physics, Korea University, Seoul 136-701, Korea}
}

\author{D.~K.~Sinclair}{
  address={HEP Division, Argonne National Laboratory, 9700 South Cass 
Avenue, Argonne, IL 60439}
}

\begin{abstract}
We compute spin-dependent matrix elements for decays of $S$-wave
quarkonia in lattice NRQCD. They appear to be in approximate agreement
with the velocity-scaling rules of NRQCD.
\end{abstract}

\maketitle


The effective field theory Nonrelativistic QCD (NRQCD)
\cite{Caswell:1985ui,Thacker:1990bm,Bodwin:1994jh} is an elegant 
formalism within which to describe heavy-quarkonium physics.
It is believed that, within NRQCD, inclusive cross sections for the 
production of a quarkonium at large transverse momentum $p_T$
can be written in a factorized form \cite{Bodwin:1994jh}:
\begin{equation}
\sigma(H)=\sum_n \sigma_n(\Lambda)
\langle 0|{\cal O}_n^H(\Lambda)|0\rangle,
\end{equation}
where the $\sigma_n$ are perturbatively calculable short-distance
coefficients, and the $\langle 0| {\cal O}_n^H(\Lambda)|0\rangle$ are
matrix elements in the vacuum state of four-fermion
operators of the form ${\cal O}_n^H=\chi^\dagger \kappa_n\psi
\left(\sum_X |H+X\rangle\langle H+X|\right) \psi^\dagger \kappa'_n\chi$.
Here, $H$ is the quarkonium state, $\psi$ is the Pauli spinor field that
annihilates a heavy quark, $\chi$ is the Pauli spinor field that creates
a heavy antiquark, and $\Lambda$ is the ultraviolet cutoff of NRQCD.
$\kappa$ contains Pauli matrices, color matrices, and the covariant
derivatives. NRQCD predicts the leading scaling behavior of the matrix
elements with $v$, the heavy-quark velocity in the quarkonium rest frame
\cite{Bodwin:1994jh}. As a consequence of these $v$-scaling rules, the
sum over operator matrix elements can be regarded as an expansion in
powers of $v$, where $v^2\approx 0.3$ for charmonium and $v^2\approx
0.1$ for bottomonium. A similar factorization formula applies to
inclusive quarkonium decays \cite{Bodwin:1994jh}, except that the matrix
elements are now between quarkonium states, rather than vacuum states,
and the four-fermion operators have the form ${\cal O}_n=\psi^\dagger
\kappa_n\chi\chi^\dagger\kappa_n'\psi$.

At large $p_T$ at the Tevatron, the dominant mechanism for $J/\psi$
production is gluon fragmentation into a $Q\overline Q$ pair, which then
evolves nonperturbatively into the $J/\psi$. The nonperturbative
evolution is described by the NRQCD matrix elements. The NRQCD
$v$-scaling rules predict that the spin-flip matrix elements are
suppressed by $v^2$ in comparison with the non-spin-flip matrix
elements. Therefore, in existing calculations, the spin-flip effects are
neglected, and the $J/\psi$ is assumed to take on the polarization of
the fragmenting gluon. Consequently, the $J/\psi$ is expected to have a
significant transverse polarization at large $p_T$
(Ref.~\cite{Cho:1994ih}). Surprisingly, the CDF data for the $J/\psi$
polarization \cite{Affolder:2000nn} show decreasing transverse
polarization with increasing $p_T$ and disagree with the NRQCD prediction
\cite{Braaten:1999qk} in the largest $p_T$ bin.

One can question whether the neglect of spin-flip processes is justified
in the case of the $J/\psi$. The $v$-scaling rules predict the leading
power of $v$ in a matrix element, but not its coefficient. It could
happen that the spin-flip matrix elements, although suppressed by $v^2$
relative to the non-spin-flip matrix elements, are not actually smaller
numerically. It has also been suggested that the $v$-scaling rules
themselves may need to be modified for charmonium
\cite{Beneke:1997av,Brambilla:1999xf,Fleming:2000ib,Sanchis-Lozano:2001rr,
Brambilla:2002nu}.

Ideally, we would settle these issues by computing the $J/\psi$
production matrix elements in a lattice QCD simulation. Unfortunately,
it is not known how to formulate the computation of production matrix
elements in lattice simulations. However, we can instead use lattice
calculations of {\it decay} matrix elements to test the validity of
using the $v$-scaling rules to estimate the sizes of matrix elements.

We take for our lattice action the discretized version of the NRQCD
action through next-to-leading order in $v$ that is given in
Ref.~\cite{Davies:1995db}. We also include improvements of
relative-order $a$ to the terms in the lattice action that are of
leading order in $v^2$, and we implement tadpole improvement
\cite{Lepage:1992xa} as described in Ref.~~\cite{Davies:1995db}. Our
initial computations were carried out on quenched gauge-field
configurations, which should reproduce the qualitative features of QCD.
Our preliminary results are based on $400$ gauge-field configurations on
$12^3\times 24$ lattices at $\beta=5.7$. We use the values $am_c=0.8$
and $am_b=3.15$ for the heavy-quark masses in lattice units. We take the
parameter $n$ that appears in the action of Ref.~\cite{Davies:1995db} to
be $1$ for bottomonium and $4$ for charmonium.

Since, at $\beta=5.7$, $a=0.81$~GeV$^{-1}$ for charmonium
\cite{Davies:1995db} and $a=0.73$~GeV$^{-1}$ for bottomonium
\cite{Davies:1994mp}, the quarkonium is well contained in the lattice
volume. However, since the quarkonium radius is approximately  $1/(mv)$,
which is about $1.2$~GeV$^{-1}$ for charmonium and about
$0.6$~GeV$^{-1}$ bottomonium, the lattice spacing is fairly coarse and
warrants the implementation of order-$a$ improvements.

Our preliminary results for $S$-wave quarkonium states are shown in 
Table~\ref{tab:me}.
\begin{table}
\begin{tabular}{c|c|c|c|c|c}
\hline
\hline
spin Transition&scaling&bottom.&bottom.\ est.
&charm.&charm.\ est.\\
\hline
singlet $\to$ triplet&$v^3/(2N_c)$&$2.72(4)\times 10^{-4}$&
$5.3\times 10^{-3}$&$2.90(3)\times 10^{-3}$&$2.7\times 10^{-2}$\\
triplet $\to$ singlet&$v^3/(2N_c)$&$9.0(1)\times 10^{-5}$&
$5.3\times 10^{-3}$&$1.13(2)\times 10^{-3}$&$2.7\times 10^{-2}$\\
singlet $\to$ singlet&$v^4/(2N_c)$&$6.5(5)\times 10^{-5}$&
$1.7\times 10^{-3}$&$9.7(2)\times 10^{-4}$&$1.5\times 10^{-2}$\\
triplet $\to$ triplet&$v^4/(2N_c)$&$6.9(6)\times 10^{-5}$&
$1.7\times 10^{-3}$&$1.02(1)\times 10^{-3}$&$1.5\times 10^{-2}$\\
triplet up $\to$ triplet up&$v^4/(2N_c)$&$6.9(6)\times 10^{-5}$&
$1.7\times 10^{-3}$&$1.02(1)\times 10^{-3}$&$1.5\times 10^{-2}$\\
triplet up $\to$ triplet long.&$v^6/(2N_c)$&$1$--$2\times 10^{-6}$&
$1.7\times 10^{-4}$&$2.8(7)\times 10^{-6}$&$4.5\times 10^{-3}$\\
triplet up $\to$ triplet down&$v^6/(2N_c)$&$<5\times 10^{-8}$&
$1.7\times 10^{-4}$&$1.4(2)\times 10^{-6}$&$4.5\times 10^{-3}$\\
\hline
\hline
\end{tabular}
\caption{Quarkonium decay matrix elements. A spin transition
labeled ``$S_i\to S_f$'' refers to the matrix element of the color-octet
$S$-wave operator of spin $S_f$ in an $S$-wave color-singlet state of
spin $S_i$. The matrix elements are normalized to the matrix element of
the ${}^3S_1$ color-singlet operator in a ${}^3S_1$ color-singlet state.
We average over unspecified spins in the quarkonium state and sum over
unspecified spins in the operators. The column labeled ``scaling'' gives
the $v$ scaling of the normalized matrix element. Here we include the
color factor $1/(2N_c)$ that arises in the free $Q\overline Q$ matrix
elements, as is suggested in Ref.~\cite{Petrelli:1997ge}. The columns
labeled ``bottom.'' and ``charm.'' give the lattice results for
the normalized matrix elements in the lowest-lying $S$-wave bottomonium
and charmonium  states. The columns labeled ``bottom.\ est.'' and
``charm.\ est.'' give the numerical values of the velocity-scaling
estimates in the ``scaling'' column, taking $v^2=0.1$ for
bottomonium and $v^2=0.3$ for charmonium. The quoted
uncertainties are statistical only. Systematic uncertainties could be
considerably larger.}
\label{tab:me}
\end{table}
As can be seen, the hierarchy of matrix elements that is predicted by
$v$-scaling is observed. However, the sizes of the matrix elements
decrease faster with increasing powers of $v$ than one would expect from
the powers of $v$ alone. The NRQCD heavy-quark spin symmetry
\cite{Bodwin:1994jh} predicts that the singlet-to-triplet transition and
the triplet-to-singlet transition should be in a ratio of 3:1, up to
corrections of order $v^2$. The lattice results agree with
this prediction. Furthermore, the ratios of bottomonium matrix elements
to charmonium matrix elements for the larger, well-measured matrix
elements agree within about a factor of two with expectations from $v$
scaling. Taken together, these results suggest that $v$ scaling is
fairly well obeyed, but that additional factors, beyond the powers $v$,
must be taken into account in order to estimate the sizes of the matrix
elements accurately.

As can be seen from Table~\ref{tab:me}, the triplet-to-singlet
transition rate is comparable to the triplet-to-triplet transition rate.
This suggests that, at large $p_T$ at the Tevatron, where the
color-octet, spin-triplet process dominates $S$-wave quarkonium
production, the $\eta_c$ production rate may be comparable to the
$J/\psi$ production rate. 

The spin-triplet up-to-longitudinal transition rate is small
compared with the spin-triplet up-to-up transition rate.
That is, the spin-flip matrix elements are suppressed, as is expected
from $v$-scaling. Hence, our preliminary results support the prediction
of large transverse polarization of $J/\psi$'s produced at large $p_T$ at
the Tevatron.

%
%
%


\begin{theacknowledgments}
Work in the High Energy Physics Division at Argonne
National Laboratory is supported by the U.~S.~Department of Energy,
Division of High Energy Physics, under Contract No.~W-31-109-ENG-38.
\end{theacknowledgments}



\bibliographystyle{aipproc}   


%


\end{document}